\documentclass [prl,twocolumn,superscriptaddress]{revtex4}

\usepackage {graphicx}
\usepackage {amsmath}

\begin{document}

\title{X-ray image reconstruction from a diffraction pattern alone}

\begin{abstract}
A solution to the inversion problem of scattering would offer 
aberration-free diffraction-limited 3D images without the resolution
 and depth-of-field limitations of lens-based tomographic systems. 
Powerful algorithms are increasingly being used to act as lenses to
 form such images.  Current image reconstruction methods, however, 
require the knowledge of the shape of the object and the low spatial 
frequencies unavoidably lost in experiments.  Diffractive imaging has 
thus previously been used to increase the resolution of images obtained
 by other means.  We demonstrate experimentally here a new inversion 
method, which reconstructs the image of the object without the need 
for any such prior knowledge.
\end{abstract}

\author{S.~Marchesini}
\email[Correspondence and requests for materials should be addressed
 to S. Marchesini: ]{smarchesini@llnl.gov}
\affiliation{Lawrence Livermore National Laboratory, 7000 East Ave., Livermore, CA 94550-9234, USA}

\author{H.~He}
\affiliation{Lawrence Berkeley National Laboratory, 1 Cyclotron Rd, Berkeley CA 94720, USA}

\author{ H.~N.~Chapman}
\author{S.~P.~Hau-Riege}
\author{ A. Noy}
\affiliation{Lawrence Livermore National Laboratory, 7000 East Ave., Livermore, CA 94550-9234, USA}

\author{ M.~R.~Howells}
\affiliation{Lawrence Berkeley National Laboratory, 1 Cyclotron Rd, Berkeley CA 94720, USA}

\author{ U.~Weierstall}
\author{ J.~C.~H.~Spence
}
\affiliation{Arizona State University, Department of Physics, Tempe AZ 85287-1504, USA}

\pacs{61.10.Nz  42.30.Rx  42.30.Wb  68.37.Yz}

\preprint{UCRL-JC-153571}

\date{\today}
\maketitle

The inversion problem of coherent scattering - the reconstruction of a 
single-scattering potential from measurements of scattered intensity in the 
far-field - has occupied physicists for over a century, and arises in fields 
as varied as optics, astronomy, X-ray crystallography, medical tomographic 
imaging, holography, electron microscopy and particle scattering generally. 
A solution to this problem would offer diffraction-limited images of 
non-periodic objects without the use or need for a lens. The 
possibility of solving the X-ray phase problem for an isolated object was 
first suggested by Sayre \cite{Sayre:1952} in 1952, who pointed out that 
Bragg diffraction in crystals undersamples the diffracted intensities 
\cite{Finer:1,Sayre:1980,Sayre:1998}. The Bragg limitation 
on sampling is lifted for non-periodic objects, which allow finer sampling 
of the diffraction pattern. An iterative phase-retrieval method, capable of 
phasing adequately sampled diffracted intensity first appeared in 1972 
\cite{Gerchberg:1972}), followed by important theoretical advances due to 
Fiddy, Bates and others---see \cite{Stark:1987} for a review. The 
iterative algorithm was greatly improved through the introduction of 
feedback and compact support (The support is the boundary of the object) by Fienup around 1982 with the 
hybrid input-output (HIO) algorithm \cite{Fienup:1982}, which allowed 
inversions of optical data \cite{Cederquist:1988,Kamura:1998}. A 
significant breakthrough occurred in 1999 with the reconstruction by Miao 
and coworkers of a two-dimensional non-periodic X-ray image at 0.075 $\mu $m 
resolution from diffraction data and a low-resolution image of the object 
\cite{Miao:1999}. Subsequent work has produced nano-crystal images at 
sub-micron resolution using hard X-rays \cite{Robinson:2001}, and 
striking tomographic images at higher resolution \cite{Miao:2002}, using 
zone-plate X-ray images to provide the low-resolution data. Images have also 
been obtained by inversion of experimental coherent electron diffraction 
patterns \cite{Weierstall:2002}, and further laser-light images 
\cite{Spence:2002}. In one proposed application, atomic-resolution images 
might be obtained by the inversion of X-ray pulses diffracted from single 
molecules \cite{Neutze:2000}. More generally, the possibility of using 
particles with low radiation damage has stimulated a burst of recent 
activity \cite{Spence:2001}.

The iterative algorithms, such as the HIO algorithm, iterate between real 
and reciprocal space, applying various \textit{a priori} constraints in
 each domain. In 
diffraction space, the modulus squared of the diffracted wavefield is set 
equal to the measured diffraction intensities, whereas in real space the 
transmission function of the object is set to zero outside the known 
boundary of the object (the support). Other constraints, which have been 
used, include the known sign of the scattering function and the various 
symmetries of the scatterer. These constraints may be classified as convex 
or non-convex, and the theory of Bregman projections can be used to 
understand their convergence properties \cite{Stark:1987}. In theory, 
and confirmed by 
simulation, the support mask need not trace the exact boundary of the object 
(loose support), and zero (transparent) regions of the object inside the 
support may converge to their correct value without being forced by the 
mask. However, in practice, a less than perfect estimate of the support 
often prevents the reconstruction of the correct image \cite{Fienup:1987}) 
--see also below. Efforts to obtain the support function of the object from 
the support of the experimentally accessible autocorrelation function have 
been proposed for special classes of objects using elegant geometrical 
methods \cite{Fienup:1983}). Methods for inverting Patterson maps to 
charge densities for crystals and non-periodic objects are reviewed in 
Buerger \cite{Buerger:1959}). Up until now, no images have been 
reconstructed from experimental x-ray diffraction patterns of arbitrary 
objects without a lower-resolution image provided by an optic.

In this article we eliminate the need for this secondary image and 
demonstrate ultrahigh resolution imaging without the need for a lens. The 
object support function is determined together with the object itself, 
without additional spatial information. The procedure builds on the HIO 
algorithm, in which constraints are iteratively applied in real and 
reciprocal space, and a feedback parameter is used in real space to damp the 
application of the support constraint. Feedback allows this algorithm to 
emerge from local minima of the invariant error metric and thus avoid 
stagnation. The uniqueness of solutions found by this method has been 
studied in detail \cite{Seldin:1990} (Rare ambiguous solutions have been found
 in two-dimensions in cases where the spectrum is factorable).
 Our innovation is the 
simple but powerful use of the current estimate of the object to determine 
the support constraint. The first estimate of the support is the support of 
the autocorrelation function \cite{Buerger:1959}. Although this estimate is far from 
accurate, it is continually updated by thresholding the intensity of a 
blurred version of the current estimate of the object under reconstruction. 
Thresholding traces the boundary of the object at a given intensity contour. 
The blurring acts to smooth out noise, and provides a form of 
regularization. In turn, through the normal behavior of the HIO algorithm, 
the improved support constraint gives rise to yet a better estimate of the 
object. We find that this method is very stable, and converges to the 
correct support and object for both simulated and experimental x-ray 
diffraction data. The algorithm also successfully reconstructs complex 
objects (those that cause large variations in the phase of the exit 
wavefield in two dimensions), which hitherto have been experimentally 
difficult to reconstruct
\cite{Cederquist:1988,Kamura:1998,Weierstall:2002}. 
This opens up the possibility of image 
reconstruction from microdiffraction patterns, where the illumination is 
tightly focused on the object.

\begin{figure*}[tbp]
\centerline{\includegraphics[width=7.1in]{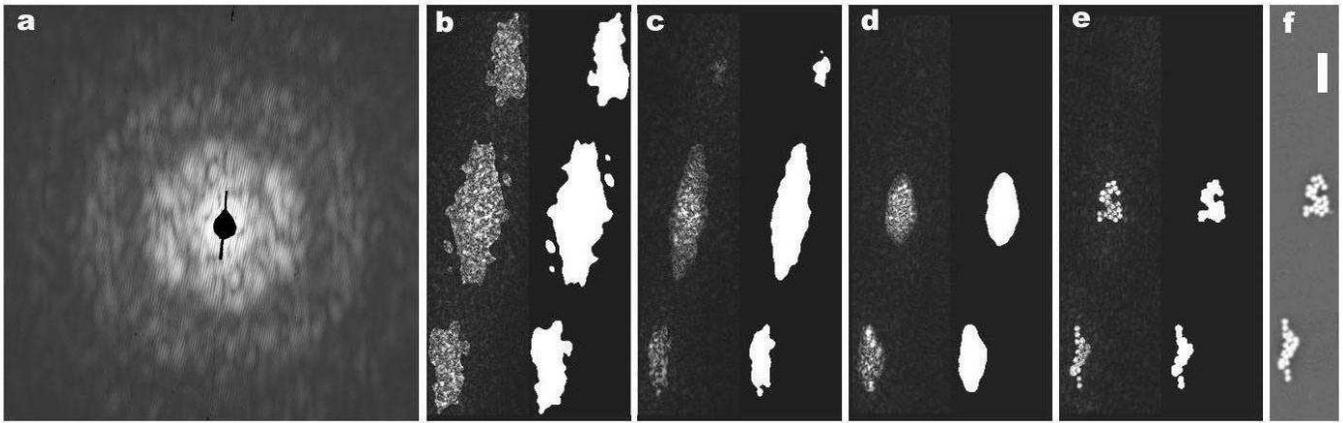}}
\caption{ Image reconstruction from an experimental X-ray diffraction 
pattern. (a) X-ray diffraction pattern of a sample of 50 nm colloidal gold 
particles, recorded at a wavelength of 2 nm. (b) to (e) shows a sequence of 
images produced by the algorithm as it converges. Number of iterations: 1 
(b), 20 (c), 100 (d), 1000 (e). The reconstruction progresses from the 
autocorrelation function in (b) to an image in (e) with a steady improvement 
of the support boundary (shown at bottom of each frame). For comparison a 
SEM micrograph of the object is shown in (f). The scale bar length is 300 nm 
and the resolution of our reconstructed image is about 20 nm.}
\label{fig1}
\end{figure*}

Details of the algorithm are as follows. We start from the autocorrelation 
function of the object. This real-space map, obtained by Fourier 
transforming the diffraction pattern, displays all "interatomic" vectors, 
with peaks for all vectors between isolated objects, shifted to a common 
origin. It contains many more peaks than the object, and, even for an 
acentric object, possesses a center of inversion symmetry. Since the object 
must fit within the autocorrelation function, our first estimate of the 
support is a mask obtained from this function using a contour at the 4{\%} 
intensity level. Both the correct object density and its centro-symmetric 
inversion fit within this initially centric mask, however inversion symmetry 
is progressively lost as the algorithm converges. We then apply the HIO 
algorithm with feedback parameter $\beta =0.9$ and the real space support 
given by the calculated mask.~We obtain the part of the diffraction pattern 
covered by a central beamstop from the transform of the current estimate of 
the object. Low frequency components are treated as free parameters. Every 
20 iterations we convolve the reconstructed image (the absolute value of the 
reconstructed wavefield) with a Gaussian of width $\sigma $ (FWHM = 2.3548 
$\sigma $ ) to find the new support mask. The mask is then obtained by 
applying a threshold at 20{\%} of its maximum. The width $\sigma $ is set to 
3 pixels in the first iteration, and reduced by 1{\%} every 20 iterations 
down to a minimum of 1.5 pixels. Similarities of the original 
Gerchberg-Saxton algorithm with the "solvent flattening" method suggest that 
this method could be extended to crystallography. 

We have tested the method using two-dimensional experimental data as well as 
two- and three-dimensional sets of simulated data. The experimental soft 
X-ray transmission diffraction pattern from two clusters of gold balls of 
50$\pm $5 nm diameter deposited on a silicon nitride window was recorded at 
the Advanced Light Source at the Lawrence Berkeley Laboratory, using soft 
x-rays soft x-rays with a wavelength of 2.1 nm 
\cite{Acta:2003,Phys:2003}. 
In Figure \ref{fig1} we present the experimental diffraction pattern and the 
sequence of images produced by the algorithm as it converges. 
As shown in the first step, the algorithm starts with a support mask with 
perfect inversion symmetry. 
After a few iterations the symmetry is broken.
First, one of the three regions of the mask disappears, and then the support 
envelope shrinks progressively around the gold ball objects.
Finally, a stable solution showing excellent agreement with a scanning 
electron microscope image of the same object is obtained. 
The solution also agrees well with a previous reconstruction by a different
 method \cite{Acta:2003}. 
Note that we would not expect a perfect match between the electron and
 x-ray images, since image formation processes are different for 
electrons and x-rays.
 Repeated computational trials have all shown the same degree of 
convergence to the correct image or its centro-symmetric inversion. 
Although after a few hundred iterations the algorithm always converged 
to the correct image (independent of the initial random choice of phases), 
as iterations were carried further both the support and the image show 
arbitrary displacements due to the translational invariance of the solution.

\begin{figure}[htbp]
\centerline{\includegraphics[width=3.4in]{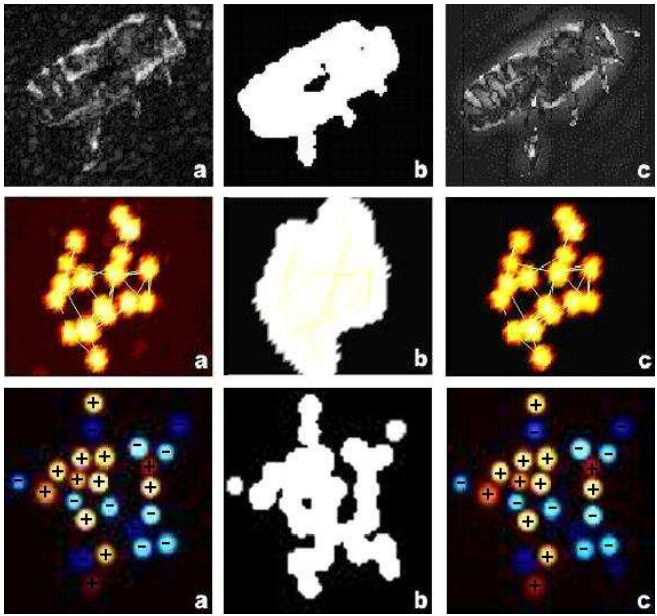}}
\caption{(Color online) Image reconstructions from simulated 
diffraction patterns of a 
gray-scale image (top row), a 3-D cluster of gold balls (center row) and
 a complex object illuminated with a complex focussed probe (bottom row). In 
each row (a) is the recovered object, (b) the recovered support and (c) the 
original image. The greyscale image demonstrates that the algorithm does not 
depend on any ``atomicity" constraint provided by the gold balls. For the 
complex object the real part is shown, blue is negative, red/yellow is 
positive.} 
\label{fig2}
\end{figure}

To further assess the validity of the algorithm we have tested it on several 
sets of simulated diffraction patterns from gold spheres and gray-scale 
images. The simulations all include noise and the loss of data due to a 
central beam-stop. They show that the algorithm is successful to the same 
degree as the standard HIO algorithm with tight support. As examples, we 
include in Figure \ref{fig2} the reconstructions of: a 
grayscale image; a 3D cluster of gold balls (ball diameter of 50$\pm $5 nm); 
and a complex object illuminated by a focused beam. The greyscale image 
demonstrates that the algorithm does not require any ``atomicity" constraint 
provided by the gold balls. The particular 3D cluster was chosen to have a 
small number of balls for visualization purposes - the algorithm also works 
with a much larger number of balls. The third example is of particular 
interest since it is well known that the reconstruction of complex objects 
is much more difficult than real objects, but is possible using either 
disjoint, precisely known or specially shaped supports 
\cite{Fienup:1987,Spence:2002}. Complex objects arise in 
optics and X-ray diffraction in two-dimensions when large phase-shifts occur 
within the eikonal approximation, or if that approximation fails, in the 
presence of spatially dependant absorption effects, and in the presence of 
multiple scattering. The question arises as to whether our new method 
provides a sufficiently tight support, especially for objects fragmented 
into separated parts, to allow the inversion of complex objects. 
Figure \ref{fig2} (bottom) shows the reconstruction of a 
cluster of gold balls where each ball is multiplied by a constant random 
phase shift between 0 and 2$\pi $. The cluster is singled-out from 
neighbouring ones by a focused beam. A perfect match between object and 
reconstruction is again observed if one takes into account the translation 
and constant phase invariance in the solution. The result is significant 
because it relaxes the requirement for plane-wave illumination. The 
generality of the technique is thus increased because now the focused probe 
can be used to isolate objects in the image field.

\begin{figure}[htbp]
\centerline{\includegraphics[width=3.4in]{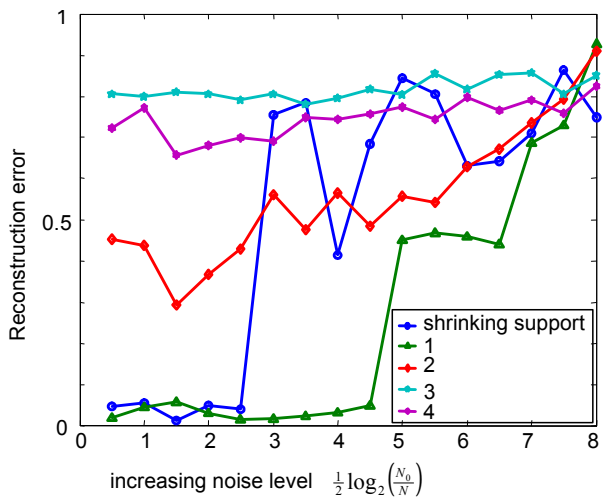}}
\caption{
(Color online) Reconstruction error of the new algorithm and the HIO 
algorithm for a complex object as a function of Poisson noise in 
the diffraction image ($N$ is the maximum number of photons 
per pixel, $N_0=2^{17}$). 
 In the HIO algorithm increasingly looser supports (support 1-4) are used:
 support 1, 2 and 3 are obtained by thresholding the original image after
 convolving with a Gaussian of 0.5, 5 and 25 pixels width. 
Support 4 is obtained from the autocorrelation. The HIO algorithm with 
perfect support (support1) works well even for high noise levels whereas 
it always fails with loose supports (supports 3,4). 
The new algorithm (shrinking support) is superior to the HIO with limited
 knowledge of the support shape (support 3,4). Our algorithm fails when
 the noise in real space becomes larger than the threshold used to update 
the support.}
\label{fig3}
\end{figure}

We have compared the behaviour of our algorithm to that of the HIO 
algorithm. The HIO algorithm requires the support \textit{a priori}, 
and as is well known the 
error in the reconstruction decreases as the support becomes tighter and 
closer to the actual boundary of the object. This is illustrated in 
Fig. \ref{fig3}, which shows plots of the reconstruction 
error, as a function of Poisson noise in the diffraction intensities, for 
the HIO algorithm with support masks of various degrees of accuracy. The 
masks for these cases were calculated by convolving the object by Gaussians 
of various widths (0.5, 5, and 25 pixels) and thresholding at 5{\%} level 
from the maximum. This corresponds to knowing the shape of the object to a 
given resolution. It is seen that even for low noise, HIO can achieve a 
reasonable reconstruction only if the support mask is set to the boundary 
known at essentially the same resolution to which we are reconstructing the 
object. The reconstruction error for our algorithm (which does not require 
\textit{a priori} knowledge of the support) is also plotted in Figure 
\ref{fig3}. We expect that the noise level at which our algorithm fails to 
reconstruct occurs when the noise in real space becomes larger than the 
threshold used to update the support. At this noise level the estimate of 
the support will be influenced by the noise, and the algorithm will be 
unable to converge to the correct boundary. This suggests that the optimum 
threshold setting depends on the noise level in the data, and we will only 
be able to reconstruct those parts of the object where the contrast is above 
the noise. As the support used in the HIO algorithm becomes looser, we 
observe our algorithm to be much superior, even in the presence of noise. 
This is because our algorithm always improves upon the support and so makes 
optimal use of the available information. The only prior knowledge needed is 
that the object possesses compact support (i.e. is isolated), so that 
oversampling diffraction conditions can be guaranteed experimentally, and 
that the contrast of the object is above the noise. By comparison with 
earlier methods \cite{Spence:2001}, no knowledge of the shape of the object is required. 
There are few adjustable parameters in our algorithm; namely, support 
resolution, support threshold, and feedback parameter. Additional 
constraints can be added to strengthen convergence, such as atomicity, 
positivity and histogram matching \cite{Elser:2003,Our:1}. 

The combination of an apparatus to measure large-angle diffraction patterns 
with our new method of data analysis forms a new type of 
diffraction-limited, aberration-free tomographic microscopy. The absence of 
inefficient optical elements makes more efficient use of damaging radiation, 
while the reconstruction from a three-dimensional diffraction data set will 
avoid the current depth-of-field limitation of zone-plate based tomography. 
The use of focused illumination will allow users to select either one or 
two-part objects (which may be complex) from a field. The conditions of beam 
energy and monochromatization used in these preliminary experiments are far 
from optimum for diffractive imaging and can be greatly improved to reduce 
recording times by more than two orders of magnitude. We expect this new 
microscopy to find many applications. Since dose scales inversely as the 
fourth power of resolution, existing measurements of damage against 
resolution can be used to show that statistically significant images of 
single cells should be obtainable by this method at 10 nm resolution in the 
0.5-10 $\mu $m thickness range under cryomicroscopy conditions. Imaging by 
harder coherent X-rays of inorganic nanostructures (such as mesoporous 
materials, aerosols and catalysts) at perhaps 2 nm resolution can be 
expected. Atomic-resolution diffractive imaging by coherent electron 
nanodiffraction has now been demonstrated \cite{Zuo:2003}. The imaging of 
dynamical systems, imaging with new radiations for which no lenses exist, 
and single molecule imaging with X-ray free-electron laser pulses remain to 
be explored.

\begin{acknowledgments}
We acknowledge stimulating discussions with Abraham Sz\"oke. 
This work was performed under the auspices of the U.S. Department of 
Energy by the Lawrence Livermore National Laboratory under Contract 
No. W-7405-ENG-48 and the Director, Office of Energy Research, 
Office of Basics Energy Sciences, Materials Sciences Division of 
the U. S. Department of Energy, under Contract No. DE-AC03-76SF00098. 
SM acknowledges funding from the National Science Foundation. 
The Center for Biophotonics, an NSF Science and Technology Center, 
is managed by the University of California, Davis, under Cooperative 
Agreement No. PHY0120999.
\end{acknowledgments}

%\appendix
%\section{Method}
%\label{appendix:method}


\begin{thebibliography}{27}
\bibitem{Sayre:1952} D.~Sayre, \textit{Acta Crystallogr.}
 \textbf{5}, 843 (1952).
\bibitem{Finer:1} Finer sampling of intensities is needed to satisfy 
Shannon's theorem (and hence solve the phase problem) since the 
autocorrelation function of the molecule acts as a bandlimit.
\bibitem{Sayre:1980} D.~Sayre, (1980). In: \textit{Image processing
 and coherence in Physics}. Springer Lecture Notes in Physics, 
Vol. 112. Eds. (M. Schlenker et al. (1980)) vol. 229.
\bibitem{Sayre:1998} D.~Sayre, H.~N.~Chapman, J.~Miao, 
\textit{Acta Crystallogr. }\textbf{A54}, 232-239 (1998).
\bibitem{Gerchberg:1972} R.~Gerchberg and W.~Saxton, \textit{Optik } 
\textbf{35}, 237 (1972).
\bibitem{Stark:1987} H.~Stark, \textit{Image Recovery: Theory and
 applications}. (Academic Press, New York, 1987).
\bibitem{Fienup:1982} J.~R.~Fienup, \textit{Appl. Opt,} \textbf{21}, 
2758 (1982).
\bibitem{Cederquist:1988} J.~N.~Cederquist, J.~R.~Fienup, 
J.~C.~Marron, R.~G.~Paxman, \textit{Opt. Lett.} \textbf{13}, 619. (1988).
\bibitem{Kamura:1998} Y.~Kamura, S.~Komatsu,
 \textit{Jpn. J. Appl. Phys. }\textbf{37}, 6018 (1998).
\bibitem{Miao:1999} J.~Miao, C.~Charalambous, 
J.~Kirz and D.~Sayre, \textit{Nature}, \textbf{400}, 342 (1999).
\bibitem{Robinson:2001} I.~K.~Robinson, I.~A.~Vartanyants,
 G.~J.Williams, M.~A.~Pfeifer, J.~A.~Pitney, 
\textit{Phys. Rev. Lett. } \textbf{87}, 195505 (2001).
\bibitem{Miao:2002} J.~Miao, T.~Ishikawa, B.~Johnson, E.~H.~Anderson,
B.~Lai, K.~O.~Hodgson, \textit{Phys. Rev. Lett.} \textbf{89}, 088303. (2002).
\bibitem{Weierstall:2002} U.~Weierstall, Q.~Chen, J.~C.~H.~Spence, 
M.~R.~Howells, M.~Isaacson, R.~R.~Panepucci, 
\textit{Ultramicroscopy} \textbf{90}, 171 (2002).
\bibitem{Spence:2002} J.~C.~H. Spence, U.~Weierstall, 
M.~R.~Howells, \textit{Philos. Trans. R. Soc. London }\textbf{360}, 875 (2002).
\bibitem{Neutze:2000}. R.~Neutze, R.~Wouts, D.~van der Spoel,
 E.~Weckert and J.~Hajdu, \textit{Nature} \textbf{406}, 752 (2000).
\bibitem{Spence:2001} J.~C.~H.~Spence, M.~R.~Howells, L. D. Marks, 
and J. Miao, \textit{Ultramicroscopy} \textbf{90}, 1 (2001).
\bibitem{Fienup:1987} J.~R.~Fienup,
 \textit{J. Opt. Soc. Am. A}\textbf{ 4}, 118 (1987).
\bibitem{Fienup:1983} J.~R.~Fienup, T.~R.~Crimmins, W.~Holsztynski, \textit{J. Opt. Soc. Am.} \textbf{72}, 610 (1982).
\bibitem{Buerger:1959} M.~J.~Buerger, 
\textit{Vector space, and its application in crystal-structure 
investigation.} New York, Wiley (1959).
\bibitem{Seldin:1990} J.~H.~Seldin, J.~R.~Fienup, 
\textit{J. Opt. Soc. Am. A}\textbf{ 7}, 412 (1990).
\bibitem{Acta:2003} H.~He, S.~Marchesini, M.~Howells, U.~Weierstall,
 G.~Hembree, J.~C.~H.~Spence, \textit{Acta Crystallogr.} \textbf{A59}, 143 (2003).
\bibitem{Phys:2003} H.~He, S. Marchesini, 
M.~Howells, U.~Weierstall, H.~Chapman, S.~Hau-Riege, A.~Noy, 
J.~C.~H.~Spence, \textit{Phys. Rev.} \textbf{B 67}  174114  (2003).
\bibitem{Elser:2003} V.~Elser, \textit{J. Opt. Soc. Am. A}
\textbf{ 20}, 40 (2003). 
\bibitem{Our:1} Our adjustment of the support provides 
precisely what the histogram constraint needs: a continually 
updated estimate of the count of pixels of low object density.
\bibitem{Zuo:2003} J.~M.~Zuo, I.~Vartanyants, M.~Gao, R.~Zhang,
 L.~A.~Nagahara, \textit{Science} 300, 1419 (2003). 
\end{thebibliography}
\end{document}